\documentclass[twocolumn]{aastex61}

\usepackage{amsmath}
\usepackage{bm}

\newcommand{\acronym}[1]{{\small{#1}}}

\newcommand{\teff}{T$_{\mathrm{eff}}$}
\newcommand{\logg}{log $g$}
\newcommand{\feh}{[Fe/H]}
\newcommand{\numax}{$\nu_{\mathrm{max}}$}
\newcommand{\dnu}{$\Delta\nu$}

\newcommand{\PS}{PS}

\newcommand{\Msun}{M$_{\odot}$}

\newcommand{\project}[1]{\textsl{#1}}
\newcommand{\gaia}{\project{Gaia}}

\newcommand{\thecannon}{\project{The~Cannon}}

\newcommand{\apogee}{\project{\acronym{APOGEE}}}

\newcommand{\lamost}{\project{\acronym{LAMOST}}}

\shorttitle{Core helium burning diagnostic with spectra }
\shortauthors{Hawkins et al.}

\begin{document}
\title{Photospheric diagnostics of core helium burning in giant stars}

\correspondingauthor{Keith Hawkins}
\email{khawkins@astro.columbia.edu}

\author[0000-0002-1423-2174]{Keith Hawkins}
\affil{Department of Astronomy, Columbia University, 550 W 120th St, New York, NY 10027, USA}

\author{Yuan-Sen Ting}
\affil{Research School of Astronomy and Astrophysics, Mount Stromlo Observatory, The Australian National University, ACT 2611, Australia}
\affil{Institute for Advanced Study, Princeton, NJ 08540, USA}
\affil{Department of Astrophysical Sciences, Princeton University, Princeton, NJ 08544, USA}
\affil{Observatories of the Carnegie Institution of Washington, Pasadena, CA 91101, USA}

\author{Hans-Walter Rix}
\affil{Max-Planck-Institut f\"ur Astronomie, K\"onigstuhl 17, D-69117 Heidelberg, Germany}

%

%
\begin{abstract}
Core helium burning primary red clump (RC) stars are evolved red giant stars which are excellent standard candles. As such, these stars are routinely used to map the Milky Way or determine the distance to other galaxies among other things. However distinguishing RC stars from their less evolved precursors, namely red giant branch (RGB) stars, is still a difficult challenge and has been deemed the domain of asteroseismology. In this letter, we use a sample of 1,676 RGB and RC stars which have both single epoch infrared spectra from the \apogee\ survey and asteroseismic parameters and classification to show that the spectra alone can be used to (1) predict asteroseismic parameters with precision high enough to (2) distinguish core helium burning RC from other giant stars with less than 2\% contamination. This will not only allow for a clean selection of a large number of standard candles across our own and other galaxies from spectroscopic surveys, but also will remove one of the primary roadblocks for stellar evolution studies of mixing and mass loss in red giant stars. 
\end{abstract}

\section{Introduction} 
\label{sec:introduction}
Red giants are evolved stars that, unlike the Sun, burn hydrogen in a shell around an inert helium core \citep{Iben1968}. If the initial mass of the star is high enough, the helium core mass grows large enough to initiate helium fusion \citep{Schwarzschild1962}. For primary core helium burning red clump (RC) stars, this happens in an abrupt event known as the helium flash. Hydrogen shell-burning red giant branch (RGB) stars and their evolutionary successors, RC stars, can appear very similar in their surface properties and spectra. Primary RC stars are excellent standard candles \citep{Stanek1998,Hawkins2017}, while RGB stars or more massive secondary RC stars of nearly the same effective temperatures (\teff) are not.Thus finding and characterizing core helium burning primary RC stars is of great importance not only for stellar evolution and Galactic archaeology but also for building a more precise cosmic distance ladder \citep{Stanek1998,Salaris2007, Bressan2013,Bovy2014,Gontcharov2017, Hawkins2017}. However, separating RC stars from less evolved shell hydrogen burning RGB stars or more massive secondary RC stars is difficult and continues to be a barrier in solving numerous problems in stellar astrophysics \citep{Bressan2013}. 

Asteroseismology has become the gold standard for distinguishing these two types of stars \citep{Montalban2010, Bedding2011, Mosser2011, Mosser2012, Stello2013, Pinsonneault2014, Vrard2016, Elsworth2017}. The solar-like oscillations in red giant stars arise from near-surface convection and can have either or both acoustic (p-mode) or gravity (g-mode) characteristics \citep{Chaplin2013}. P-modes are primarily associated with the stellar envelope with pressure as a restoring force, while g-modes probe its core with buoyancy as a restoring force. The observed stellar pulsations, which mostly contain pure p-modes, have been studied using the frequency power spectrum of well-sampled photometric light curves of nearby stars. For evolved stars, there is coupling between g- and p-modes. However, for RC stars the core density is lower than that of RGB stars of a similar luminosity and radius, which causes a significantly stronger coupling between the g-mode and p-modes leading to the appearance of observable `mixed modes' in the oscillations spectrum. 

These features include the large frequency separation (\dnu), defined as the frequency between adjacent p-modes with the same angular degree ($\ell$) but different radial order ($n$) and the frequency at which there is maximum power (\numax). For more evolved stars showing a mixed mode pattern, the period separation (\PS) between the mixed modes can also be measured. The asteroseismic scaling relations \citep{Kjeldsen1995,Chaplin2013} relate theses features to different stellar properties. Specifically, \numax\ has been shown to relate to the surface gravity (\logg) and \teff\ such that \numax~$\propto g\sqrt{\mathrm{T_{eff} }}$, while \dnu\ is proportional to the square root of the mean stellar density and thereby $\sqrt{(g/R)}$. Chiefly, for the case of evolved stars, the period separation \PS\ has been shown to distinguish RGB stars, with low \PS, from RC stars, with high \PS\ \citep{Montalban2010, Bedding2011,Vrard2016}. 

Photospheric diagnostics, such as spectroscopy, for distinguishing between RC and RGB stars have largely been overlooked. However, stellar evolutionary isochrones indicate that RGB and RC stars likely have a different distribution in \teff-\logg\ space \citep{Bovy2014}. In addition, there is extra mixing along the red giant branch and even potentially at the He flash which separates RGB and RC stars in their C and N  abundances \citep{Martell2008, Masseron2015, Masseron2017, Masseron2017b}. Therefore, we posit that a star's spectrum can be used to discern whether it is burning helium in its core and predict, indirectly, its asteroseismic parameters.
\section{Data}
We explore here stars which contain both seismic information, namely \PS\ and \dnu\ from\ \cite{Vrard2016} and moderate resolution (R$\sim$22,000) $H$-band spectra, our photospheric probe, from the thirteenth data release of the Apache Point Observatory Galactic Evolution Experiment survey \citep[APOGEE,][]{Holtzman2015,Majewski2015}.  We then made a quality cuts, keeping only those for which both the \apogee\ STARFLAG and ASPCAP flag were set to zero meaning they are reliable and the uncertainty in \PS\ was less than 10~s. This was to remove stars where the spectra had obvious problems or the \PS\ was poorly constrained. The latter is necessary as \PS\ is the asteroseismic parameter that best distinguishes between RGB and RC stars. These cuts reduced the final sample to 1,676 stars which have high-quality \apogee\ spectra and precisely measured \teff, \feh, \dnu, and \PS. 
 
\begin{figure}[h]
\centering
\includegraphics[width=1.0\columnwidth]{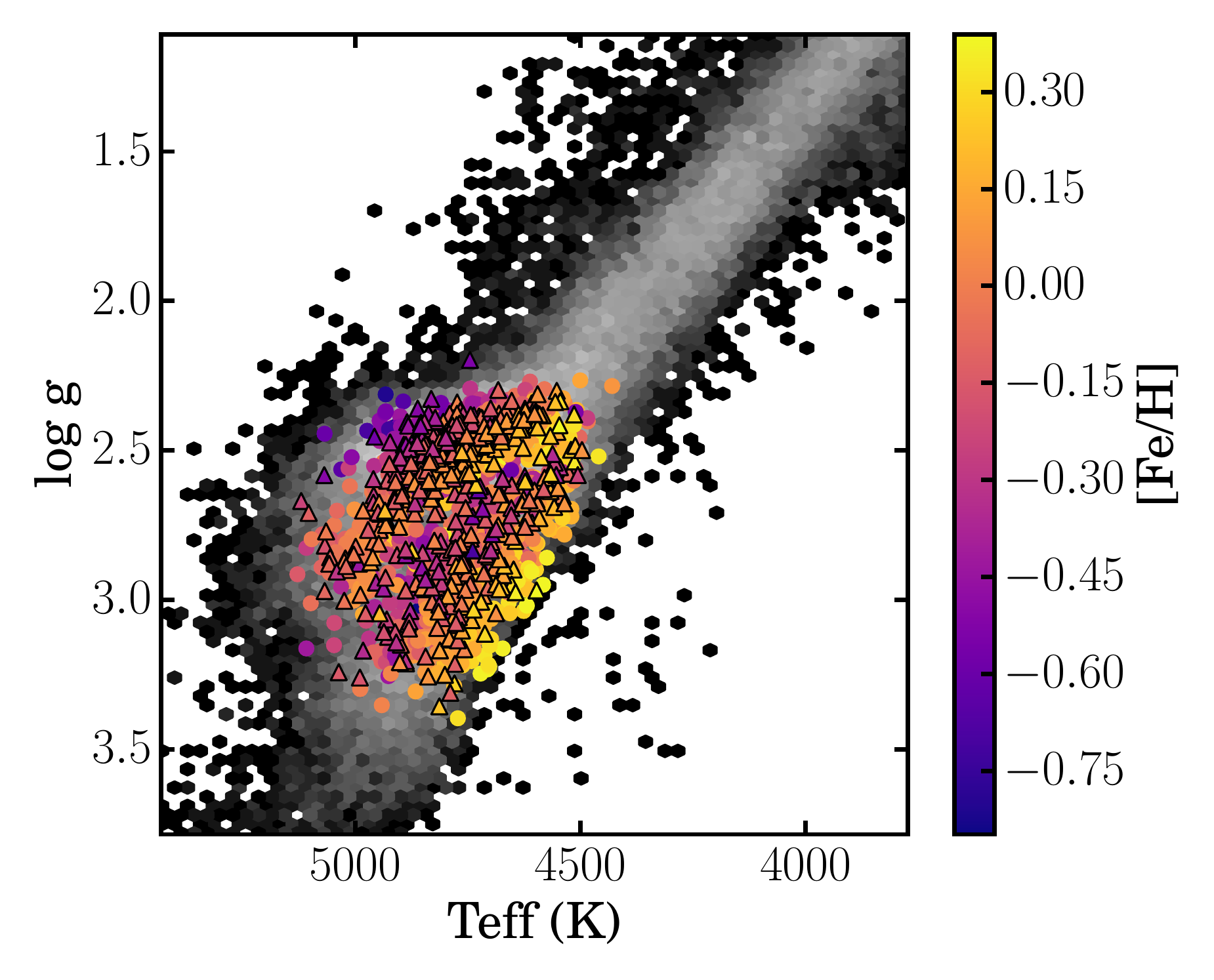}
\caption{The spectroscopic HRD (i.e. \logg\ as a function of \teff) for the training set (circles) and the test set (triangles). Each are color-coded by the star's \feh. The gray log density scale background shows the full \apogee\ DR13 sample. } 
\label{fig:HRD}
\end{figure}

According to asteroseismic classification, there are 576~RGB and 1100~RC stars in our final sample. We note that 219 of the RC stars are classified as non-standard candle secondary RC \citep{Girardi1999}, which are more massive and slightly less luminous. Typical uncertainties are 70~K, 0.04~dex, 0.002~$\mu$Hz, and 2.81~s for \teff, \feh, \dnu, and \PS, respectively. The sample was randomly divided into a training set which contained 70\% of stars (1,173 stars), and a test set which contained the remaining 30\% (503 stars). In Fig.~\ref{fig:HRD} we show the spectroscopic Hertzsprung-Russell diagram (HRD) of the training set (circles) and the test set (triangles). 

\section{Results and Discussion}
\subsection{Distinguishing red clump stars using predicted asteroseismic parameters}
For our work, we made used \thecannon\ \citep{Ness2015, Casey2016}, in order to ascertain if the spectra contained the information to not only predict the asteroseismic values (i.e. \dnu\ and \PS), but also distinguish those stars which are burning helium in their cores. \thecannon\ is a data-driven algorithm that uses a generative model of stellar spectra by predicting the flux in each pixel as a polynomial function of the stellar and asteroseismic parameters (i.e.  \teff, \feh, \dnu, and \PS). Our setup of \thecannon\ models the flux at each wavelength as a quadratic polynomial of \teff, \feh, \dnu, and \PS. Our model does not currently include \numax\ as a label, however since \numax\ and \dnu\ are strongly correlated \citep{Chaplin2013, Yang2012}, predicting one will constrain the other. During the training step, the stellar and asteroseismic labels are fixed, and  the coefficients of the polynomial that best reproduce the spectra are determined. In the test step, the coefficients found in the training set are fixed, while the stellar and asteroseismic labels are determined for the test set stars. This last step is specifically done for cross validation purposes. For a detailed introduction to \thecannon, the algorithm, and setup for \apogee, we refer the reader the release papers \citep{Ness2015, Casey2016}. 
\begin{figure}
\includegraphics[width=1.05\columnwidth]{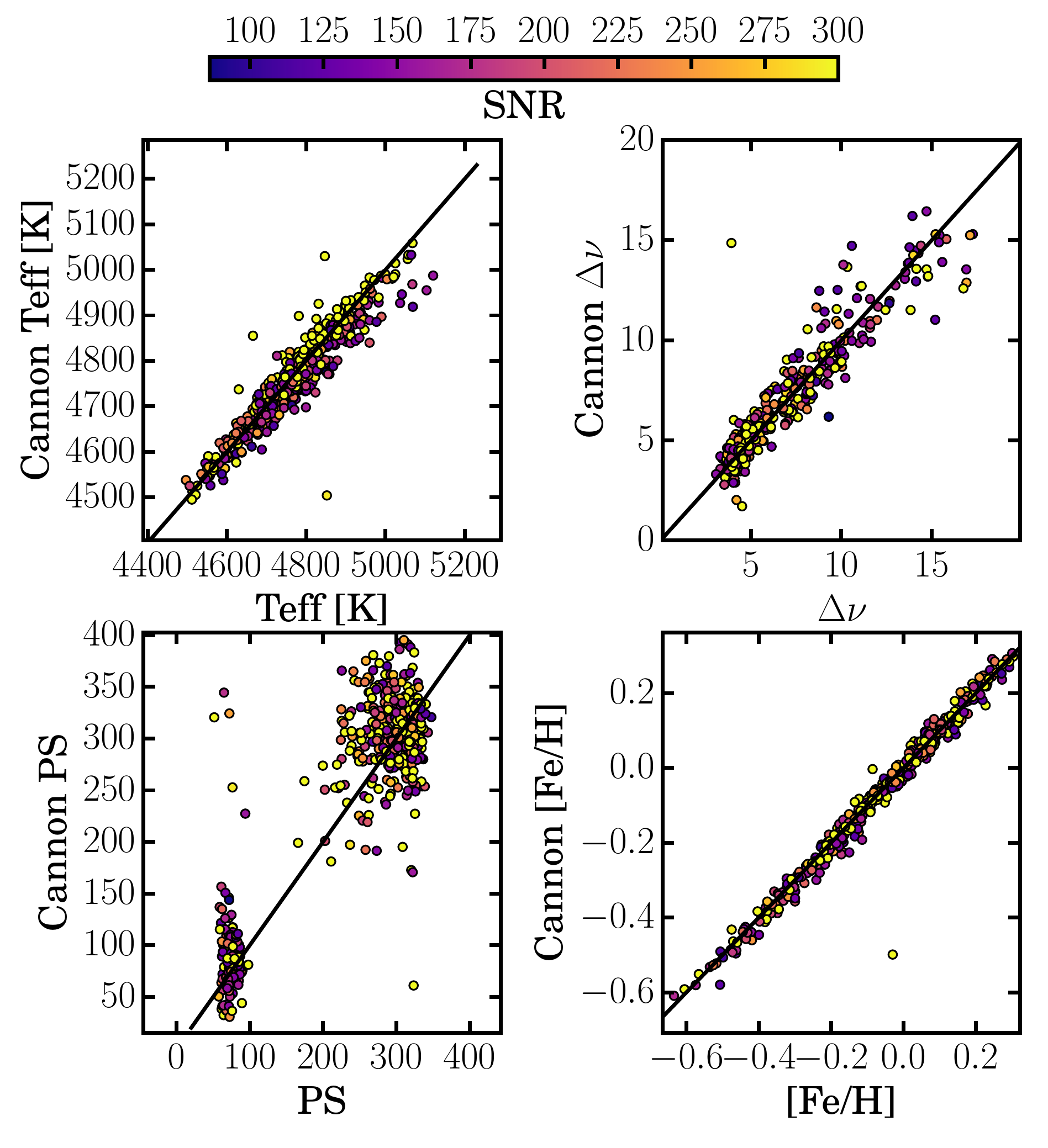}
\caption{The performance of \thecannon\ with respect to the known \teff\ (top left), \dnu\  (top right), \PS\ (bottom left), and \feh\ (bottom right) for stars in the test set. The black lines in each panel represents the 1:1 line. The known \teff\ and \feh\ are taken from \apogee\ DR13 while the known \dnu\ and \PS\ are taken from \cite{Vrard2016}. The color coding represent the signal-to-noise raito of the spectra. }
\label{fig:performance}
\end{figure}

The results for the test set were able to reproduce \teff\ with a bias of less than 10~K and dispersion of 37~K, \dnu\ with a bias of less than 0.01~$\mu$Hz and dispersion of 1.23~$\mu$Hz, \PS\ with a bias of less than --8~s and dispersion of 47~s, and \feh\ with no bias and dispersion of 0.02~dex  (Figure~\ref{fig:performance}).  As illustrated in Figure~\ref{fig:performance}, each of the parameters largely follow the one-to-one relationship indicating good agreement between the values predicted by \thecannon\ and the established ones. Since the signal-to-noise ratio (SNR) of our sample is high (larger than SNR $\geq$ 100), there does not appear to be a loss of performance with decreasing SNR. 

\begin{figure*}
\includegraphics[width=2.\columnwidth]{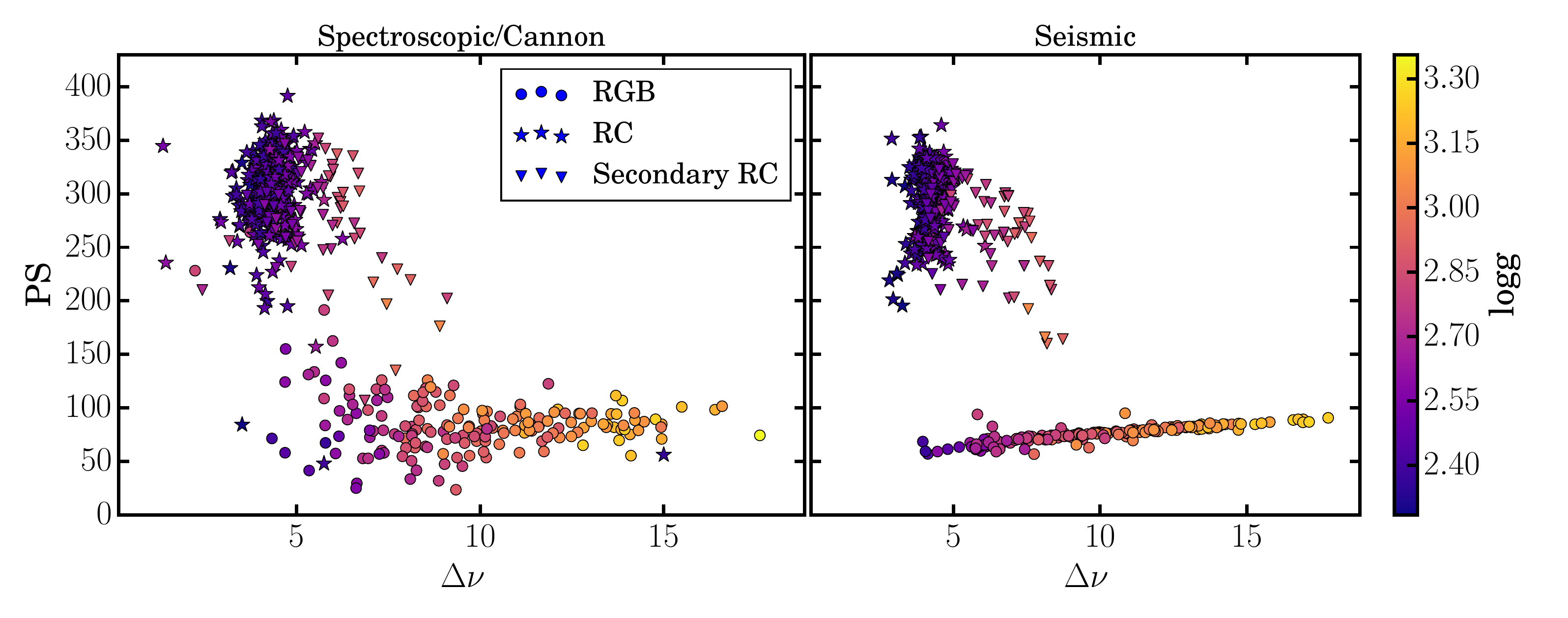}
\caption{The \PS\ as a function of \dnu\ measured using asteroseismology (right) and predicted by spectroscopy (left) for RC stars (stars), secondary RC stars (triangles), and RGB stars (circles). For the both left and right panels only the values for the test set are shown. Each are color-coded by the \apogee\ measured \logg. The two visible populations in both spectroscopy and asteroseismology indicate either can be used to distinguish RC and RGB stars from one another. }
\label{fig:PS}
\end{figure*}

The right panel of Fig.~\ref{fig:PS} shows the \PS\ as a function of \dnu\ for RGB (circles), RC (stars), and the more massive secondary RC (triangles) for test set stars as determined by asteroseismology \citep{Vrard2016}. The points are color-coded by the spectroscopic \logg. It is clear that RC stars are separated from RGB stars in this plane. The left panel of Fig.~2 shows the spectroscopically predicted values of \PS\ and \dnu\ for the same stars. Fig.~\ref{fig:PS} indicates that \apogee\ spectra contain enough information to not only predict several asteroseismic parameters but also whether the core of a red giant star is burning helium or inert as well as. We however note that while this result represents a first step to distinguishing between RGB and RC stars in a robust and purely spectroscopic way, the signatures seen in this work are currently restricted to the Kepler and CoRoT fields. More data and further tests are required to show that it works for populations with different underlying metallicity or mass distributions than these fields.

Additionally, it may be possible to separate primary and secondary RC stars from each other using the predicted \dnu\ parameter \citep{Yang2012}, though further study will be required. The RC false positive rate of our method (i.e. the number of non-RC stars in the test set which are falsely classified as RC stars) is $\sim$2\% globally but depends on the the location in the HRD (see section~\ref{subsec:contam} for more details). For reference, the contamination of non-RC stars  thought to be $\sim$9\% globally for other spectro-photometric methods \citep{Bovy2014}. Fig.~\ref{fig:PS} also illustrates that \dnu\ and \logg\ are correlated, consistent with the asteroseismic scaling relations \citep{Kjeldsen1995}. 
 
\subsection{What spectral features predict core helium burning? }
\begin{figure*}
\includegraphics[width=2.\columnwidth]{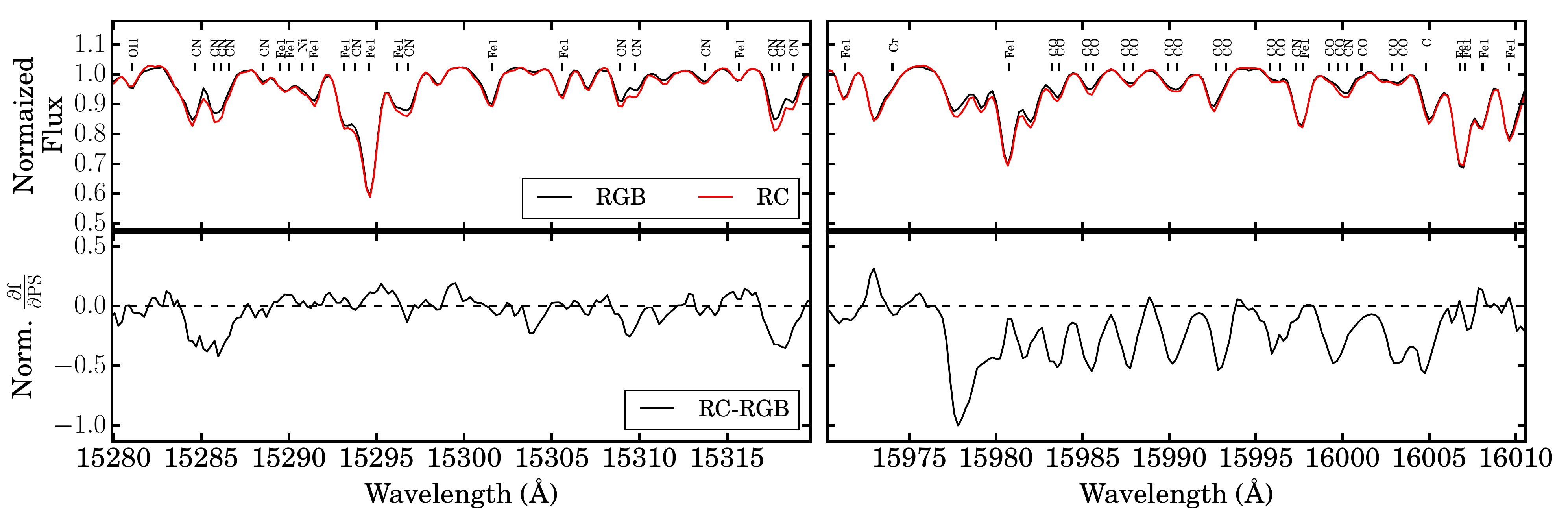}
\caption{We compare stacked spectra of 5~RGB (black line) and 7~RC (red line) stars which are found in the same part of the H-R diagram (i.e. similar \teff, seismic \logg, and \feh). The spectral regions (one centered at 15300~\AA\ on the left and one centered at 15990~\AA\ on the right) were chosen based on those where the spectral model predicts that it contains the most information to separate RGB and RC stars. Line identifications for atomic and molecular features taken from an Arcturus atlas\citep{Hinkle2005} are also shown. Upon controlling for the stellar atmospheric parameters, core helium burning RC stars are distinct around CN, C, and CO spectral lines. The bottom left and right panels illustrates the normalized partial derivative of the flux with respect to \PS\ at fixed \teff, \dnu, and \feh\ predicted by \thecannon. The most sensitive spectral features to \PS\ have a normalized flux partial derivative of --1 or 1.   }
\label{fig:spec}
\end{figure*}

One particular strength of \thecannon\ is that it can be used to discover the spectral regions which are most sensitive to a particular stellar or asteroseismic label. In Fig.~\ref{fig:spec} we show a median stacked spectrum of 5~RGB (black) and 7~RC (red) stars which have the same stellar parameters (i.e. the \teff, \logg, and \feh\ are, within the uncertainties, equivalent) in two spectral regions which are sensitive to \PS\ (the asteroseismic label which most distinguishes RC and RGB stars). The stacked spectra of RGB and RC stars with the same stellar parameters are remarkably similar except for specific features.

Fig.~\ref{fig:spec}, which shows two such spectral region between 15280 -- 15325~\AA\ (on the left) and 15970 -- 16010~\AA\ (on the right), indicates the differences between RGB and RC stars lie mostly around molecular CN and CO~line features \citep{Hinkle2005}. The differences in these molecular features may be expected because there is thought to be extra mixing along the red giant branch from the red giant branch bump to the helium flash which would make RC stars lower in their carbon to nitrogen ratio compared to RGB stars of the same stellar parameter \citep[e.g.][]{Martell2008, Lagarde2012, Masseron2015, Masseron2017, Masseron2017b, Hawkins2016b}. 

\begin{figure}
\includegraphics[width=1.0\columnwidth]{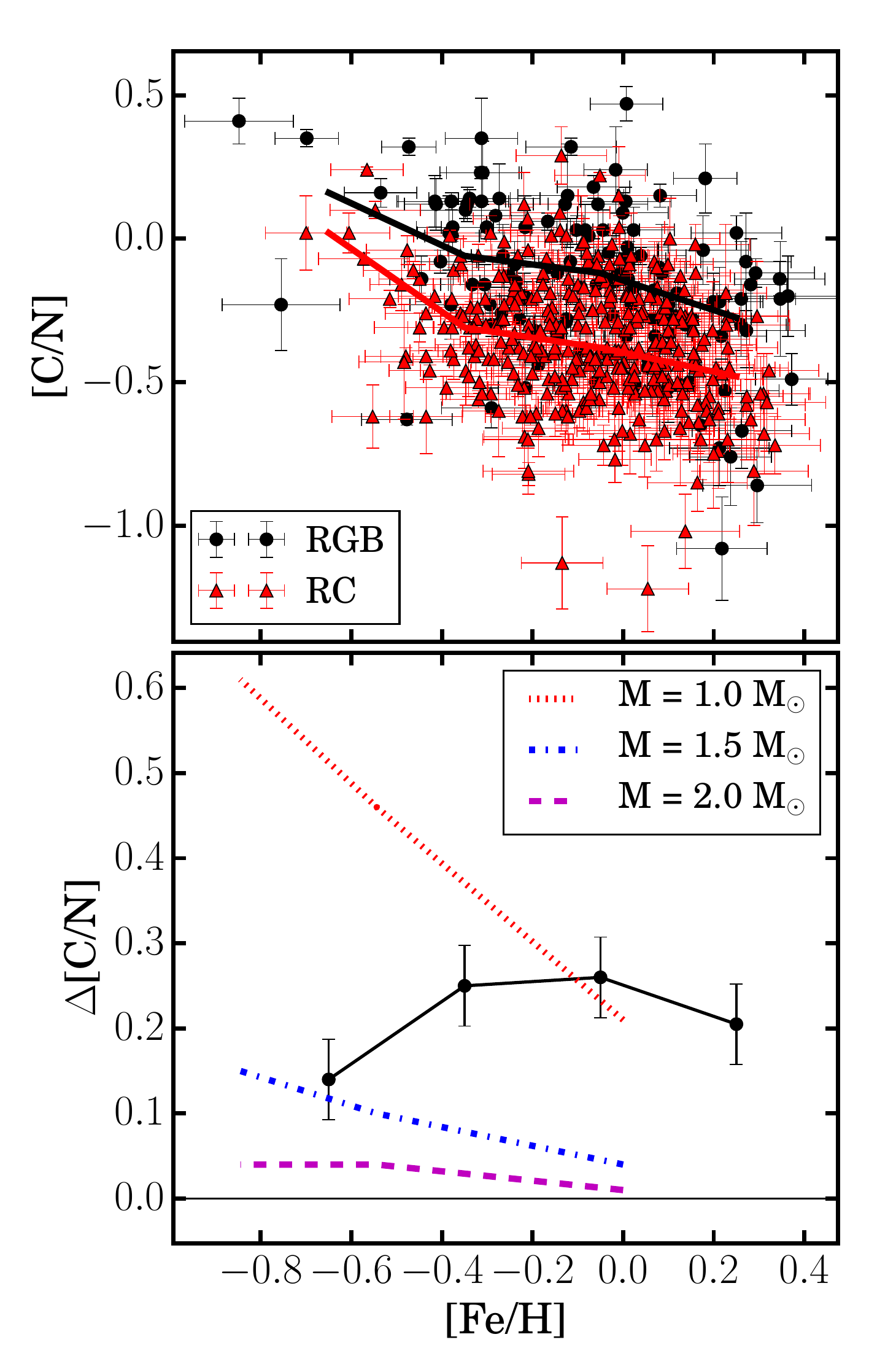}
\caption{Top Panel: Illustrates the [C/N] as a function of \feh\ for RGB (black circles) and RC (red triangles) which are in common with our sample and that of \cite{Hawkins2016b}, respectively (see the text for further explanation on the data used). The solid black, red, and cyan lines represent the running median of the RGB, RC and secondary RC in this space, respectively. Bottom Panel: Illustrates the difference in [C/N], $\Delta$[C/N], between RGB and RC stars (black line) as a function of \feh. Also shown are the predicted difference in the surface [C/N] abundance ratio between RGB and RC stars as a function of metallicity using the models of a 1~\Msun\ (red dotted line), 1.5~\Msun\ (blue dash-dotted line), and  2.0~\Msun\ (magenta dashed line) star from \cite{Lagarde2012}.}
\label{fig:CNdiff}
\end{figure}

An example of this can be found in Fig.~1 of \cite{Masseron2015}. These authors draw on the grid of stellar models from \cite{Lagarde2012} and showed that for a star with an initial mass of  1~\Msun\ at solar metallicity the expected difference in the surface [C/N] abundance ratio between a RGB and RC star, denoted as $\Delta$[C/N], is $\sim$0.20~dex (with the RC star having a lower [C/N] ratio compared to a RGB star of similar surface gravity). Additionally, in that work they also show that the difference in the  [C/N] ratio  between RGB and RC stars is both mass and metallicity dependent. Fig.~1 of \cite{Masseron2015} illustrates that it can be as high as $\Delta$[C/N]$\sim$ 0.50~dex (for M~=~1~\Msun\ and [Fe/H]~=~--0.50) down to just $\Delta$[C/N]$\sim$ 0.01~dex (for M~=~2~\Msun\ and [Fe/H]~=~0.0).

Our result in Fig.~\ref{fig:spec} seems to tell a consistent story, that C and N are useful parameters to help distinguish between RGB and RC stars. Therefore in the top panel of Fig.~\ref{fig:CNdiff}, we plot the measured surface [C/N] abundance ratio for RGB (black circles) and RC (red triangles) stars for the final sample. The running median [C/N] value as a function of \feh\ are shown as solid lines. The [C/N] ratios are taken from \cite{Hawkins2016b} as opposed to \apogee\ DR13. This was done because there are known issues in the C (and potentially N) abundances in \apogee. These issues are likely due to a discrepancy in spectroscopic and seismic \logg\ between RGB and RC stars  \citep[e.g.][]{Masseron2017b}. Since \cite{Hawkins2016b} is one of the only catalogues which derives the chemical abundances in a consistent way using the precise and accurate seismically determined \logg\ information, it is preferred over the \apogee\ values. Though similar differences in the [C/N] ratio are also found for the DR13 values.

The bottom panel of  Fig.~\ref{fig:CNdiff} shows the difference in [C/N] between RGB and RC stars (black line) and RGB in four bins. There is a clear difference such that RC stars are $\sim$0.20~dex lower in [C/N] compared to RGB stars over the range in full metallicities. This is consistent with the  theoretical expectations for stars with initial masses between 1--2~\Msun\ using the grid of model from \cite{Lagarde2012} with thermohaline convection and rotation-induced mixing included. In addition, the secondary RC stars have as much as a 0.4~dex difference in their [C/N] surface ratios compared to RGB stars of similar \logg. This is consistent with the theoretical expectations for stars with initial masses between 2--3~\Msun\ using the grid of model from \cite{Lagarde2012}. Fig.~\ref{fig:spec} and Fig.~\ref{fig:CNdiff} together indicate that there is likely extra mixing along the RGB which can cause slight surface abundance changes in C and N, which allow us to distinguish between core helium burning RC and shell hydrogen burning RGB stars that overlap on the HRD.  


Future work will be needed to further understand the exact physical processes by which each spectral feature responsible for the distinction between RGB and RC stars is observable.

\subsection{Contamination} \label{subsec:contam}
The selection of pure core helium burning RC stars is of great importance for Galactic archeology, stellar evolution, and the cosmic distance ladder as a whole. Therefore, we address here the contamination level that is expected from both RGB and secondary RC in different regions of the HRD for the method presented in this work. We also compare it to the rates found using other methods. For reference, the contamination level is measured as the false positive rate (FPR). 

In order to address the level of contamination, we made use of a sample of asteroseismically classified red giant stars \cite{Pinsonneault2014, Elsworth2017} which are not found in our test or training sets. All of these stars, while having both \apogee\ spectra and an asteroseismic classification, do not have publicly available \PS\ and/or \dnu\ information.  This secondary test sample is excluded in the study outlined in the main text because the goal of this work is two-fold: not only we aim to separate RGB from RC (which can be done with the secondary test sample), but also we also explore whether the asteroseismic parameters can be predicted from single epoch spectroscopy. This secondary test set contains 621 stars. We choose not to use the original test set because in this supplementary section, we aim to quantify the FPR for different parts of HRD which is not possible with the original test set due to its somewhat limited coverage.

To complete a exploration of the contamination level, we re-trained \thecannon\ using the same setup described above for the full 1,676 stars in our main sample. For classification purposes, we built a 3-component Gaussian mixture model in the \dnu-\PS\ plane using the asteroseismic values and classification of the full main sample. A 3-component mixture was chosen to classify stars into RGB, RC, and secondary RC populations. Additionally,   while the secondary RC is core helium burning, they are not standard candles and thus are consider contaminates for the purposes of this analysis.  We predicted the \teff, \feh, \dnu, and \PS\ for the secondary test set of 621 stars and inferred a probability for each star belonging to the RC using the mixture model. RC stars are classified as those which have a probability of belonging to the primary RC population greater than 95\% (or less than a 5\% probability of belonging to either the secondary RC or RGB population).
\begin{figure}
\includegraphics[width=\columnwidth]{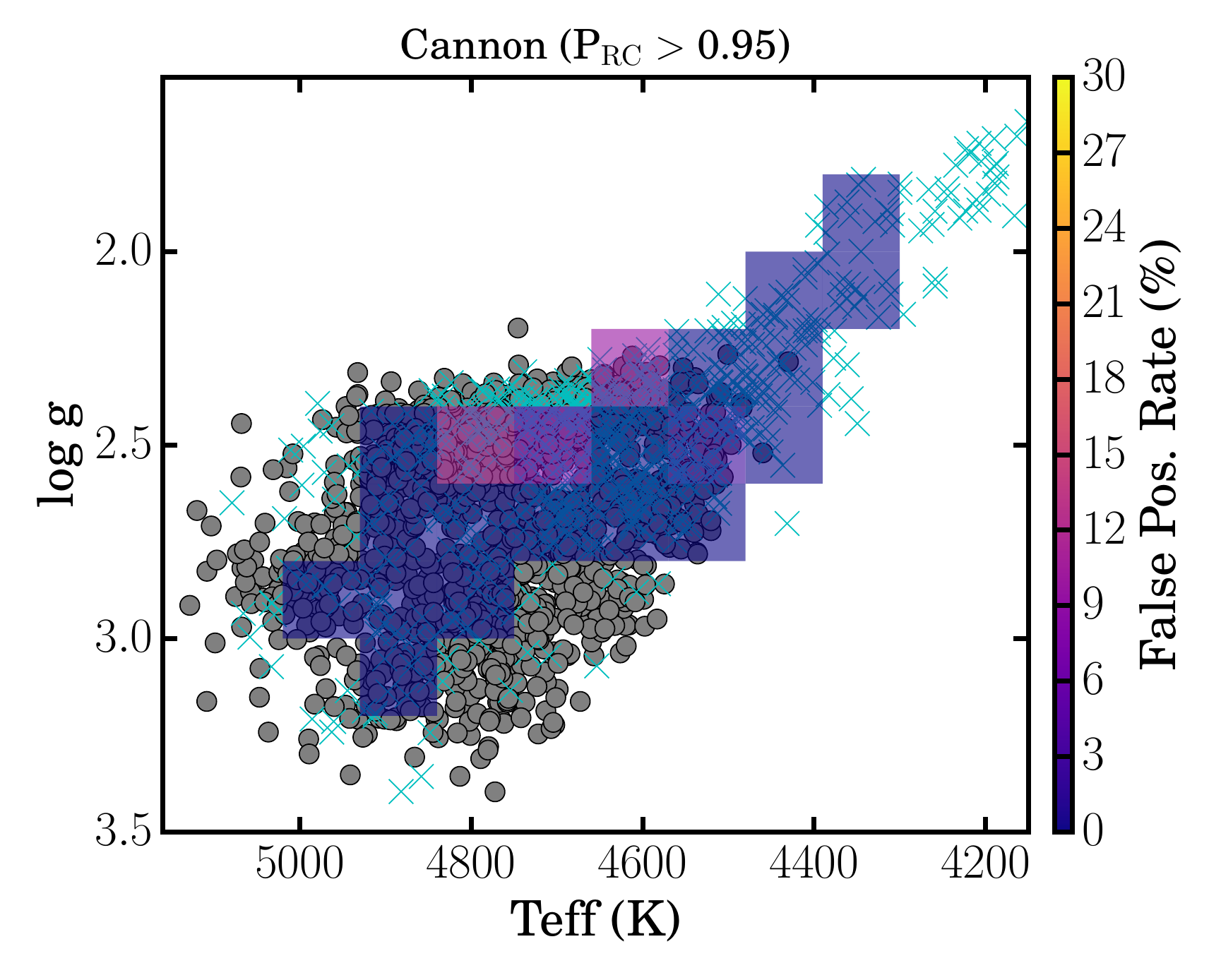}
\caption{The \logg-\teff\ diagram for the main sample (gray circles) and secondary test set (cyan x's). The false positive rate for RC stars classified using our method for specified \teff-\logg\ bins are represented by the color.  
 } 
\label{fig:FPRus}
\end{figure}

\begin{figure}
\includegraphics[width=\columnwidth]{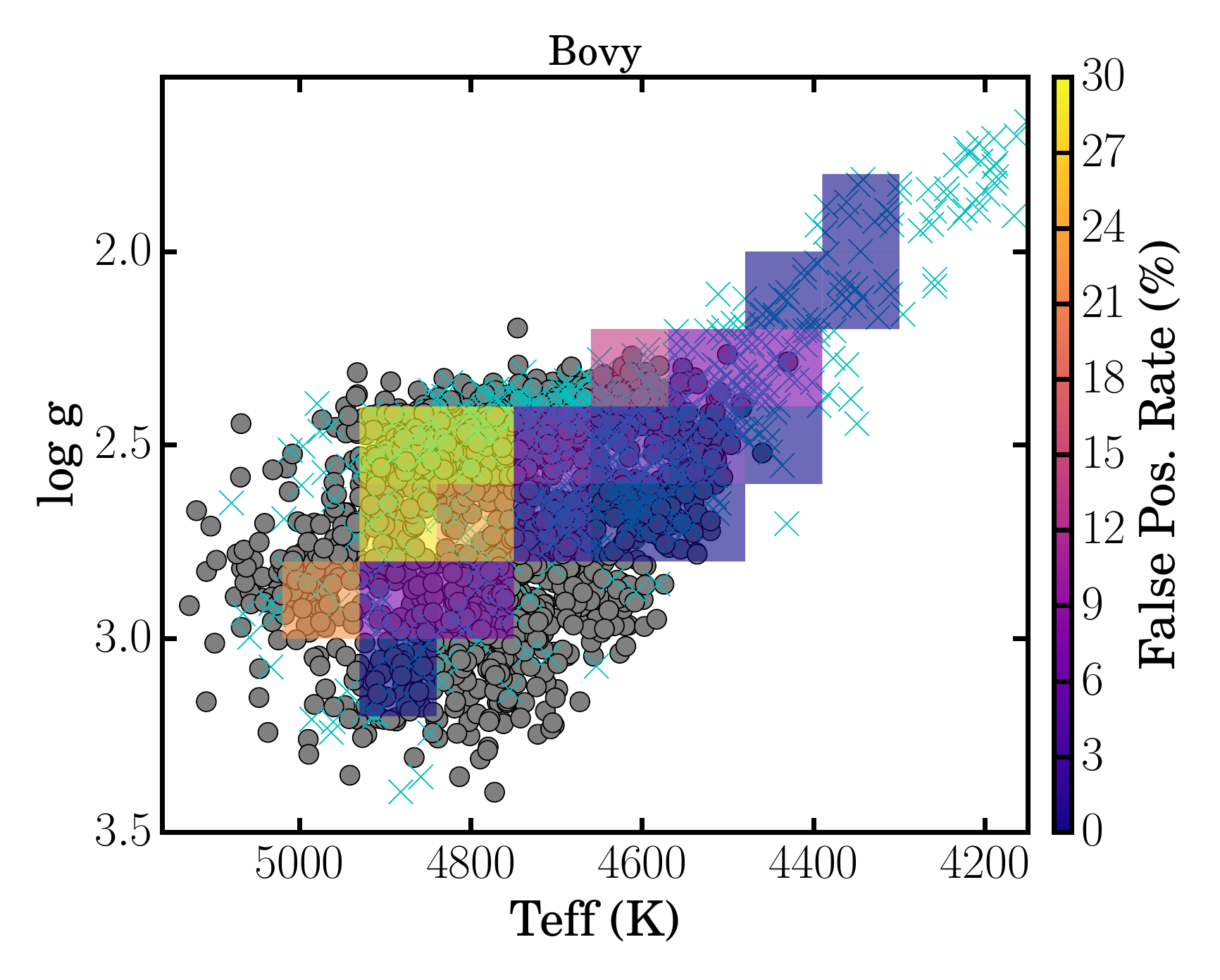}
\caption{The same as Fig.~\ref{fig:FPRus} but for RC stars classified using the method presented in \cite{Bovy2014}. }
\label{fig:FPRBovy}
\end{figure}

In Fig.~\ref{fig:FPRus} we show the spectroscopic HRD for the full 1,676 stars in the main sample (gray circles) and secondary test set (cyan x's). The colored boxes in Fig.~\ref{fig:FPRus} represent the false positive rate in \teff-\logg\ bins where there are at least 5 (non)RC stars for our method. For comparison, in Fig.~\ref{fig:FPRBovy} we illustrate the RC false positive rate in different parts of the spectroscopic HRD for the same stars but classified using the simple \teff, \logg, \feh, $(J-K_{s})$ cuts described in \cite{Bovy2014}. 

The overall false positive rate for our method is less than 2\% while it is significantly larger ($\sim$9\%) for the other spectroscopic methods presented in \cite{Bovy2014}. We remind the reader that this result is based on the assumption that the asteroseismic classification is the `ground truth'. However, as shown in Fig.~\ref{fig:FPRus} and Fig.~\ref{fig:FPRus}, the actual false positive rate is worse in regions where there is contamination from either the secondary clump or RGB. Our method largely improves the overall false positive rate of traditional spectroscopic cuts especially in regions of the HRD where there is significant contamination from other non-RC stars. These tests indicate the new method presented in the work will ultimately allow for a clean selection (at the few percent level or better) of core helium burning RC stars. One way to improve the situation and further reduce the false positive rate is to add the exquisite parallax information from the upcoming release of the \gaia\ spacecraft. The addition of this information will allow for strong constraints on the star's surface gravity (and thereby density), which can help improve the RGB-RC distinction.


\section{Summary}
In this paper, we present a robust way to determine whether a red giant star is undergoing core helium burning (RC stars) or shell hydrogen burning (RGB stars) using single epoch spectroscopy. While this spectroscopic method is more indirect than the gold standard of asteroseismology, it enables an increase in our ability to generate large and clean samples of RC stars because there many more stars with spectra than seismic information. Ultimately, the ability to cleanly separate RGB and RC stars from spectroscopy can be applied not only to \apogee\ but other large surveys such as the \lamost\ \citep{Xiang2017} survey, which contains more than five million low-resolution optical spectra, to achieve exceedingly precise distances to many stars enabling detailed studies the structure of the Galaxy. An upcoming study will test the wider applicability of the method outlined here by presenting a catalogue of RC stars in both the field and clusters from LAMOST and APOGEE (Ting, Hawkins, Rix, in prep.). An additional important application to these newly found photospheric probes of core helium burning is to aid in producing input catalogues for future asteroseismic missions. 

\acknowledgements
We thank the anonymous referee, M. Asplund, and J. Bovy for helpful comments that improved this work. K.H. is funded by the Simons Foundation Society of Fellows and the Flatiron Institute Center for Computational Astrophysics in New York City.  Y.S.T is supported by the Australian Research Council Discovery Program DP160103747, the Carnegie-Princeton Fellowship and the Martin A. and Helen Chooljian Membership from the Institute for Advanced Study at Princeton.  This project was developed in part at the 2017 Heidelberg Gaia Sprint, hosted by the Max-Planck-Institut for Astronomie, Heidelberg. This project has made use of  \software{The Cannon \citep{Ness2015, Casey2016},} which can be found at \url{https://github.com/andycasey/AnniesLasso/archive/master.zip}.\\
 Funding for the Sloan Digital Sky Survey IV has been provided by the Alfred P. Sloan Foundation, the U.S. Department of Energy Office of Science, and the Participating Institutions. SDSS-IV acknowledges
support and resources from the Center for High-Performance Computing at
the University of Utah. The SDSS web site is www.sdss.org.

SDSS-IV is managed by the Astrophysical Research Consortium for the 
Participating Institutions of the SDSS Collaboration including the 
Brazilian Participation Group, the Carnegie Institution for Science, 
Carnegie Mellon University, the Chilean Participation Group, the French Participation Group, Harvard-Smithsonian Center for Astrophysics, 
Instituto de Astrof\'isica de Canarias, The Johns Hopkins University, 
Kavli Institute for the Physics and Mathematics of the Universe (IPMU) / 
University of Tokyo, Lawrence Berkeley National Laboratory, 
Leibniz Institut f\"ur Astrophysik Potsdam (AIP),  
Max-Planck-Institut f\"ur Astronomie (MPIA Heidelberg), 
Max-Planck-Institut f\"ur Astrophysik (MPA Garching), 
Max-Planck-Institut f\"ur Extraterrestrische Physik (MPE), 
National Astronomical Observatories of China, New Mexico State University, 
New York University, University of Notre Dame, 
Observat\'ario Nacional / MCTI, The Ohio State University, 
Pennsylvania State University, Shanghai Astronomical Observatory, 
United Kingdom Participation Group,
Universidad Nacional Aut\'onoma de M\'exico, University of Arizona, 
University of Colorado Boulder, University of Oxford, University of Portsmouth, 
University of Utah, University of Virginia, University of Washington, University of Wisconsin, 
Vanderbilt University, and Yale University.


\end{document}